\documentclass[12pt,letterpaper]{article}
\usepackage[top=1.4in,bottom=1.4in,left=0.80in,right=0.80in]{geometry}
\usepackage{graphicx}

\usepackage{xcolor}
\usepackage{amsmath}
\usepackage{comment}
\usepackage{authblk}
\usepackage[super,comma,compress]{natbib}

\date{}
\begin{document}
\title{Nanoscale Surfactant Transport: Bridging Molecular and Continuum Models}

\author{Muhammad Rizwanur Rahman${^{1,*}}$, James P. Ewen ${^1}$, Li Shen${^1}$, D. M. Heyes${^1}$, Daniele Dini${^1}$, and E. R. Smith${^2}$}
\affil{${^1}$Department of Mechanical Engineering, Imperial College London, South Kensington Campus, London SW7 2AZ, United Kingdom \newline ${^2}$Department of Mechanical and Aerospace Engineering, Brunel University London, Uxbridge, London UB8 3PH, United Kingdom \newline ${^*}$ Email: m.rahman20@imperial.ac.uk}

\maketitle

\begin{abstract}
Surfactant transport is central to a diverse range of natural phenomena, and for many practical applications in physics and engineering. Surprisingly, this process remains relatively poorly understood at the molecular scale. This study investigates the mechanism behind the transport of surfactant monolayers on flat and curved liquid-vapor interfaces using non-equilibrium molecular dynamics simulations, which are compared with the continuum transport model. This approach not only provides fresh molecular-level insight into surfactant dynamics, but also confirms the nanoscale mechanism of the lateral migration of surfactant molecules along a thin film that continuously deforms as surfactants spread. By connecting the continuum model—where the long wave approximations prevail, to the molecular details—where such approximations break down, we establish that the transport equation preserves substantial accuracy in capturing the underlying physics. Moreover, the relative importance of the different mechanisms of the transport process are identified. Consequently, we derive a novel, exact molecular equation for surfactant transport along a deforming surface. Finally, our findings demonstrate that the spreading of surfactants at the molecular scale adheres to expected scaling laws and aligns well with experimental observations.
\end{abstract}

\section{Introduction} 
The dynamics of surfactants at the liquid-vapor interface represents a critical area of study with widespread implications across various disciplines, including engineering and medicine. This is underscored by a significant body of research highlighting its importance in processes ranging from respiration \citep{stetten2018surfactant,possmayer2023pulmonary} to breaking waves \citep{erinin2023effects}, and to solar energy applications \citep{morciano2020solar}. 
The spread of surfactants, and their subsequent reorganization and reorientation, lead to gradients in concentration, causing differential surface tensions within the surface plane. Such gradients result in the development of Marangoni stresses \citep{scriven1960marangoni}, thus influencing the dynamics in complex ways \citep{basu2007shaping, roche2014marangoni, trittel2019marangoni, benouaguef2021solutocapillary}. 

The lateral migration of surfactant molecules across the liquid-vapor interface involves the formation of a precursor film that alters surface tension in its path \citep{afsar2003spreading} $-$ an observation exemplifying the interwoven nature of transport. 
The implicit role played by the collective behavior of the surfactant molecules has led to them being referred to as `hidden variables' controlling fluid flows \citep{manikantan2020surfactant}.
\citet{stone1990simple} derived a simple yet versatile convection-diffusion equation, building on the form of \citet{scriven1960dynamics}, to describe the transport of surfactant molecules along a deforming liquid-vapor interface. 
\citet{gaver1990dynamics} showed that the initial distribution of surface tension, which directly depends on the surfactant concentration, dictates the system dynamics. 
Later, \citet{crowdy2021viscous} showed that the complex Burgers equation can be employed to describe surfactant transport, which was further developed through exact solutions \citep{bickel2022exact,crowdy2023fast}. However, the case-specific nature of these solutions hinders the establishment of a broader understanding of the underlying physical principles \citep{temprano2024self}.

The continuum framework, with its reliance on long-wave approximations \citep{Craster2009}, often falls short in addressing the subtleties at the nanoscale $-$ a domain uniquely suited to the strengths of a molecular dynamics simulations \citep{Evans1979}.
Specifically, the dynamic nature of the surfactant transport process makes non-equilibrium molecular dynamics (NEMD) a particularly appropriate tool to tackle these problems.
Nevertheless, nanoscale processes which govern surfactant transport along the interface have remained relatively under-explored \citep{maldarelli2022continuum}. 
Previous MD simulations of surfactants at the interfaces have mostly focused on the lateral diffusion \citep{rideg2012lateral}, surface-bulk surfactant partitions \citep{Yang2014,kandu2024interface}, and surface tension calculations \citep{Sresht2017,Peng2021}.
Of the few studies that looked at the dynamical behavior of surfactant molecules at interfaces, \citet{johansson2022molecular} studied a system of two immiscible Lennard-Jones fluids with surfactants at the interface. Their study highlighted the importance of local surfactant concentration and interfacial viscosity in accurately modeling the  flows. 

To our knowledge, there is an absence in the literature concerning the lateral transport mechanism of surfactant molecules along a deforming surface. 
This is a gap which this study aims to fill by modeling the transport of SDS surfactants on the liquid-vapor surface of a thin water film by employing NEMD simulations, and simultaneously solving the transport equation. 
An objective of this work is to establish the extent to which continuum transport equation can model the redistribution of surfactant molecules, their consequences, and what aspects of the model require molecular description.

\section{Problem description and methodology}
\begin{figure}[!t]
    \centering
    \includegraphics[width=.95\linewidth]{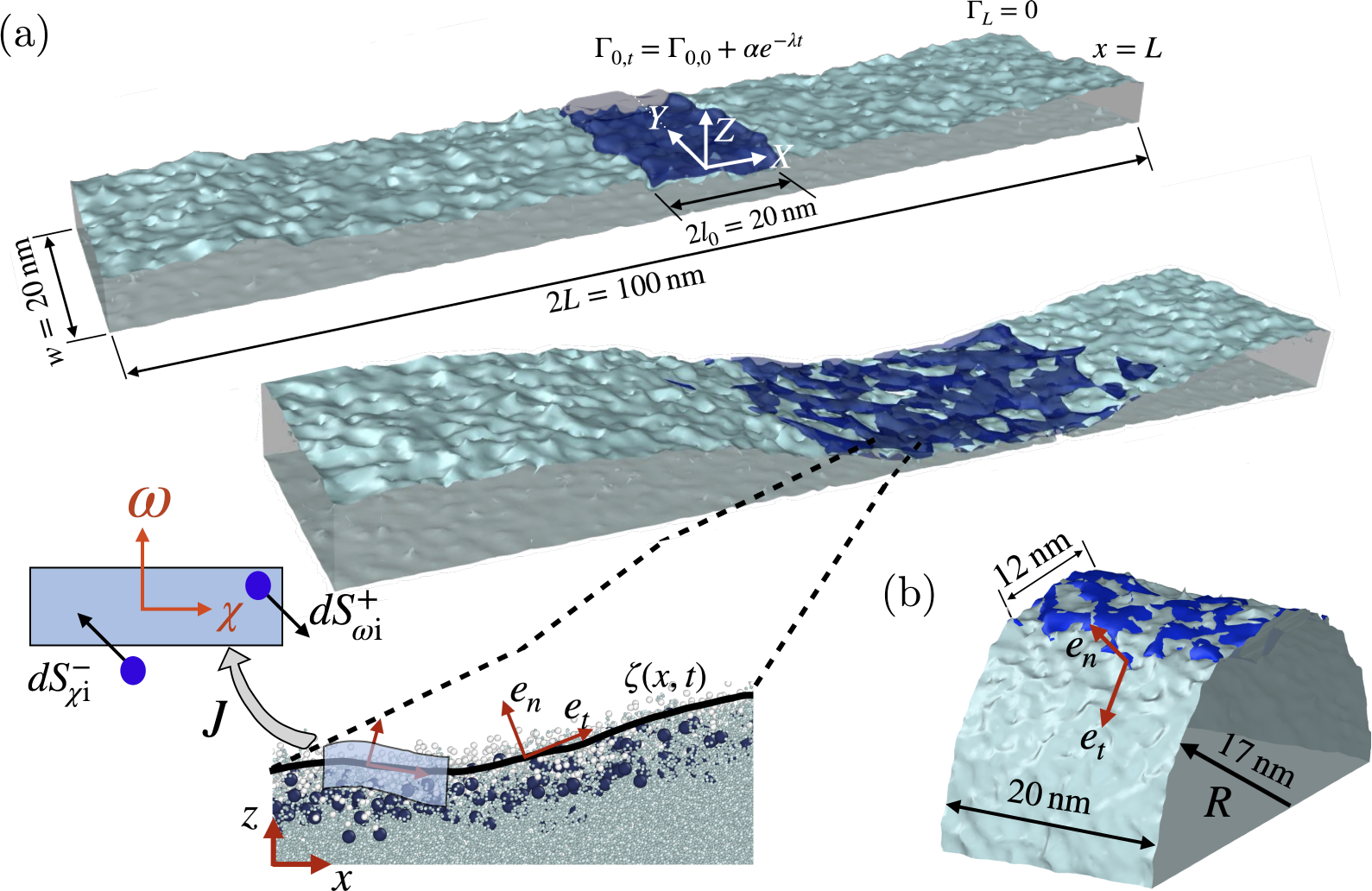}
    \caption{(a) A monolayer of sodium dodecyl sulfate deposited on the central area of a thin water film of dimensions $100 \mathrm{nm} \times 20 \mathrm{nm} \times 10 \mathrm{nm}$, where the third dimension denotes film thickness, $h_0$. Top-panel: surfactant initially concentrated over an area of $20 \mathrm{nm} \times 20 \mathrm{nm}$.
    Bottom panel illustrates the spreading of the monolayer across the surface, and subsequent deformation of the film.
    The local  curvature of the surfactant-water-air interface is shown in the zoomed inset, the surface element $d\mathbf{s}$ is resolved into normal, $\mathbf{e_n}$ and tangential, $\mathbf{e_t}$ components. Solid line shows spline fitting, $\zeta(x,t)$ to the surface. The mapping from $(x, y, z)$ space to $(\xi, \psi, \omega)$ space through Jacobean is schematically illustrated. (b) Surfactant transport along the surface of a (cylindrical) droplet of radius, $R\sim17\,\mathrm{nm}$, and width, $w\sim 20$ nm. Initially the surfactant molecules occupied an area of $\sim 12\,\text{nm}\times w$ at both the top and bottom surfaces. In both a and b, only the upper halves of the film/drop are shown.
    }
    \label{fig:shcematic}
\end{figure}

A thin water film of length, $2L=100\,\mathrm{nm}$, width, $w=20\,\mathrm{nm}$, and thickness $h=10\,\mathrm{nm}$ is considered. A monolayer of sodium dodecyl sulfate (SDS) is placed at the central area of both the top and bottom surfaces of the film which spans an area of $20\times20\, \mathrm{nm^2}$, see Fig. \ref{fig:shcematic}\,(a - top panel).
A droplet simulation is shown in panel b in Fig.\ref{fig:shcematic}, $R=17\,\mathrm{nm}$, $w=20\,\mathrm{nm}$, effective length $2L=2\pi R \approx 106.8\, \mathrm{nm}$ with similar volume to the film, carried out with a surfactant monolayer added to a point on the surface. 
The presence of surfactant reduces the local surface tension as compared to the surfactant free regions.
  
\subsection{Molecular description of the system}
A coarse-grained modeling approach is employed. The MARTINI polarizable water model \citep{yesylevskyy2010polarizable} is used here, which considers both the Lennard-Jones interactions with surrounding particles, and Coulombic interactions.
the SDS molecule is described by a five-bead coarse-grained model \citep{weiand2023effects}.  
The force field parameters, as well as the coarse grain descriptions are detailed in the \textit{Supplementary Material}.
The film (or, the droplet) and the surfactant monolayers are separately energy minimized, thermostated (at $300$~K) with the Nos{\'e}-Hoover thermostat \citep{nose1984molecular,hoover1985canonical}, and are sufficiently pre-equilibrated (for $5\,\mathrm{ns}$).
This approach leads to a well equilibrated initial state before commencing the production phase within the canonical (NVT) ensemble. Simulations are carried out using the open source code LAMMPS. A velocity-Verlet integrator is used with a timestep of 5 fs. All properties from MD simulations are averaged over the width of the film (y-axis) for improved statistics. 

\subsection{Continuum description of surfactant transport}\label{sec:transport_eq}
The spreading of the surfactant layer across the liquid-air interface, results in a locally dilute surfactant-water layer that extends over a larger surface area. This dynamic behavior of the surfactant monolayer is described by the advection-diffusion equation \citep{stone1990simple,gaver1990dynamics} $-$ a continuum model that elegantly captures the underlying physical mechanisms of the transport phenomena on the surface:
\begin{equation}\label{eq:cd}
 \underbrace{\partial_t \Gamma}_{\mathrm{unsteady}} + \underbrace{ \nabla_s \cdot (\Gamma \mathbf{u_s}) }_\mathrm{advection} + \underbrace{\Gamma (\nabla_s \cdot \mathbf{e_n})(\mathbf{u} \cdot \mathbf{e_n})}_{\mathrm{geometric\,effect}} = \underbrace{D_s \nabla_s^2 \Gamma}_\mathrm{diffusion} \,\, +\underbrace{j_n}_\mathrm{source\, term}
\end{equation}
Here, $\Gamma$ denotes the local surfactant concentration, $\mathbf{u_s}$ represents the velocity vector along the surface, $\mathbf{e_n}$ is the unit normal vector to the surface, $\nabla_s = (\mathbf{I-e_n e_n})$ is the surface gradient operator and $D_s$ characterizes the surfactant diffusion rate across the interface. 
The second term in (\ref{eq:cd}) describe surfactant advection along the surface and $D_s \nabla_s^2 \Gamma$ elucidates the diffusive transport of surfactants. The local adsorption or desorption of surfactants is modeled by $j_n$, which acts as a source or sink in the system.
Through this term, surfactants are introduced to, or eliminated from the interface, influencing the overall dynamics of the transport process; however, we have avoided this term by carefully selecting the surface concentrations, $\Gamma_0$ to obtain $j_n \to 0$.  
An implicit finite difference scheme is used to solve the transport equation, (\ref{eq:cd}), which employs both spatial and temporal discretizations to capture the dynamics of transport due to advection, diffusion, and surface curvature effects. 
The advection term is discretized using upwinding \citep{hirsch2007numerical}, whereas, a second order central differencing is applied for the diffusion term. These have been further detailed in the \textit{Supplementary Material}.

\subsection{Reconciliation of MD with continuum framework}\label{sec:reconciliation}

Any attempt to trace the surfactant particles in experimental measurements inevitably alters their intrinsic properties, thereby affecting the system's dynamics \citep{manikantan2020surfactant}. 
Conversely, the molecular system under investigation here affords direct access to individual molecules and their instantaneous properties, providing an ideal platform to verify the efficacy of (\ref{eq:cd}) in describing the transport phenomena. 
The second and third term of (\ref{eq:cd}) captures the effects of surface compression and expansion due to the non-uniformity of advective flows and curvature. These terms together account for the redistribution of surfactants driven by surface dynamics, including the Marangoni effect. 
The subscript `s' signifies that the quantities should be considered along the tangential direction to the surface. Specially, the third term $-$ which is a product of local curvature, ($\kappa=\nabla_s \cdot \mathbf{e_n}$) and the surface normal velocity, ($v=\mathbf{u} \cdot \mathbf{e_n}$) $-$ highlights the geometric effect of the locally deforming interface on the transport process. 
In quantifying these contributions in the transport of the surfactants, the SDS molecules at the deforming interface were tracked over time. 

Tracking surfactant molecules in the MD model allows us to propose a molecular version of the transport equation as:
\begin{equation}\label{eq:molecular}
 \! \!\!\!\!   \underbrace{\sum_{i=1}^{N_S} \frac{1}{J_i} \frac{d}{dt} \bigg[ m_i J_i \delta(\tilde{\boldsymbol{r}} - \tilde{\boldsymbol{r}}_i)   \bigg]}_{\mathrm{unsteady}} 
    + \underbrace{ \frac{\partial }{\partial \boldsymbol{s}} \cdot  \sum_{i=1}^{N_S} m_i   \dot{\boldsymbol{s}}_i  \delta(\tilde{\boldsymbol{r}} - \tilde{\boldsymbol{r}}_i) \vphantom{\frac{1}{J_i}}}_{\mathrm{advection}}  - \underbrace{ \sum_{i=1}^{N_S} m_i  \frac{\dot{J}_i}{J_i} \delta(\tilde{\boldsymbol{r}} - \tilde{\boldsymbol{r}}_i) }_{\mathrm{geometric\, effect}} =   \underbrace{je \vphantom{\sum_{i=1}^{N_S}}}_{\mathrm{source\, term}} 
\end{equation}  
Here, $\tilde{\boldsymbol{r}} = [\chi, \psi, \omega]$ with $\chi$, $\psi$ and $\omega$ mapped coordinates following, respectively, the two tangential and normal directions to the interface $\zeta (x,y,z,t)$. The sum is over all surfactant molecules $N_S$, and the Dirac delta function, $\delta$, includes only molecule when their position $\tilde{\boldsymbol{r}}_i$ is located at $\tilde{\boldsymbol{r}}$, with the mass $m_i$ and velocities along the surface $\dot{\boldsymbol{s}}_i = [\dot{\chi}_i, \dot{\psi}_i]$. The Jacobian $J_i$ is evaluated at the position of particle $i$ and surface normal velocity $\dot{\omega}_i$ appears in the source term (a full derivation from the control volume form is provided in the \textit{Supplementary Material}).
The under braces identify the equivalence between the terms in (\ref{eq:molecular}) and those in the continuum equation (\ref{eq:cd}). 
The unsteady term represents the temporal change of molecules within a control volume. The advection term corresponds to the flux of molecules along the tangential direction of the deforming surface.
The source term in (\ref{eq:molecular}) describes the flux of surfactant from the bulk to the surface, under the assumption that no surfactant molecules leave the surface. 
The geometric effect term, characterized by $\dot{J}_i/J_i$, accounts for the impact of the deforming volume over time, but expressed only at the location of the molecules.

There is no obvious diffusion contribution, of the form introduced by \citet{stone1990simple}, obtained in the molecular system, which is consistent with \citet{scriven1960dynamics}. 
It might be possible to justify a diffusion like effect through either $i)$ consideration of Ficks law style flux, $ii)$ interactions with the non-surfactant molecules or $iii)$ through fluctuations, which remains when the molecular equations are averaged over time, although this is not considered further here.
The presented equation (\ref{eq:molecular}) is formally exact, and simply counts the number of molecules moving into and out of a surface tracking (Lagrangian) volume, with a term $\dot{J}_i/J_i$ for the deformation of the volume in time.

A careful tracking of the spatially and temporarily evolving surface to get the local properties is necessary to attain an accurate molecular description of the transport process. 
First, a functional form of the surface, $\zeta (x,t)$ is obtained by fitting a spline to the liquid-vapor interface. Then the curvature at each point on the surface is  obtained using the derivatives, i.e.,
$\kappa(x) = {\zeta''}/{\left[1 + \zeta'^2\right]^{3/2}}$,
where $\zeta'$ and  $\zeta''$  represent the first and second derivatives of the spline fit with respect to $x$, respectively. Notably, any derivative at the molecular scale is noisy, hence the Savitzky-Golay filter was used which smooths the data without significantly distorting the signal's higher moments.
These calculations enable finding the spatio-temporal surface vectors, i.e., $\mathbf{e_t}$ and $\mathbf{e_n}$, which are instrumental in resolving the velocities of surfactant molecules along, and normal to the interface. 

Surfactant molecules on a curved surface can occupy a larger area compared to a flat surface, potentially slowing down the evolution of the concentration gradient. Additionally, the paths available for diffusion are geometrically constrained by curvature, leading to anisotropic diffusion \citep{shu2001curvature,ramirez2021effects}. 
In a dedicated series of simulations focusing on uniformly distributed surfactants on a water film, the surface diffusion of SDS molecules was estimated. To distinguish surface diffusion from bulk diffusion, only the surfactant molecules that remained on the film's surface were tracked. The trajectories of these molecules were analyzed to calculate the mean squared displacement (MSD) over the simulation period, i.e.,  $\mathrm{MSD} = \langle |s(t_0-t) - s(t_0)|^2 \rangle$, where $s(t)$ is the position of a molecule at time $t$ and $s(t_0)$ is its initial position. 
Subsequently, we employed Einstein's equation for surface diffusion, $D_s = \mathrm{MSD}/t$ \citep{rideg2012lateral}, see \textit{Supplementary Material} for further details. The diffusion coefficient  obtained shows a concentration dependence, i.e., $D_s = 3.22\mbox{e}^{-4N/3}$, where $N$ is the number of SDS molecules per unit area. This functional relation employed in solving (\ref{eq:cd}) $-$ which originally considered a constant diffusion coefficient, accounts for variations in local diffusivity caused by the varying local surfactant concentration.

\begin{figure} [!t]
    \centering
    \includegraphics[width=0.98\linewidth]{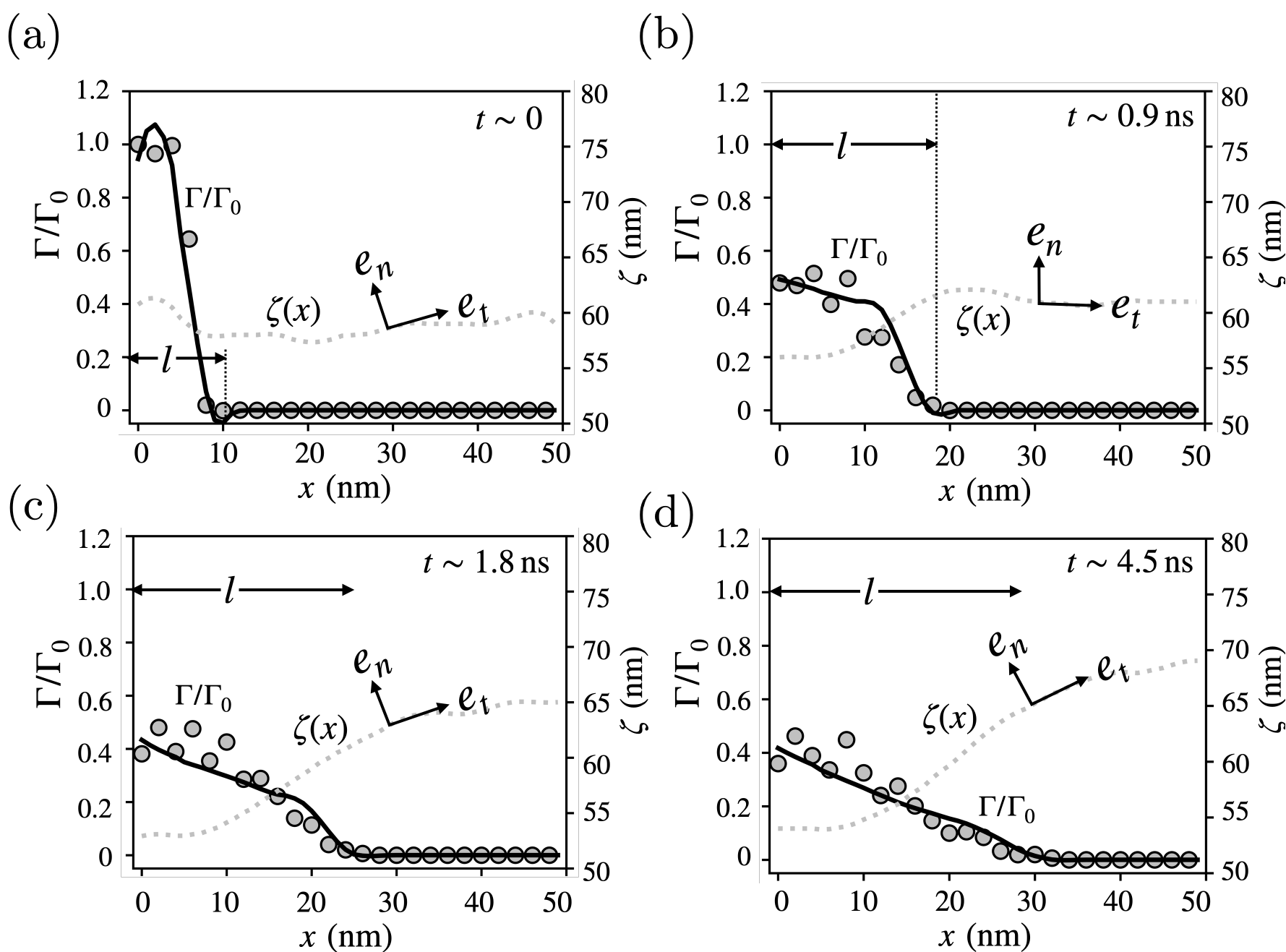}
    \caption{Spatio-temporal variation of normalized surfactant concentration along the film's length at distinct time instances. 
    $x=0$ denotes the center of the film. Solid line represents surfactant evolution captured by the continuum transport equation (\ref{eq:cd}), while symbols denote data from molecular simulations. The dotted line corresponds to spline fitting to the evolving surface. The arrows show the local tangential ($e_t$) and normal ($e_n$) surface vectors (not to scale), $l$ denotes the spreading length. 
    }
    \label{fig:compare-concentration}
\end{figure}
\vspace{-4mm}
\section{Results and Discussions}
\subsection{Surfactant transport}

Figure \ref{fig:compare-concentration} examines the time-evolution of the local surfactant concentration, $\Gamma$ (normalized by the initial concentration at $x=0$, i.e., $\Gamma_0$) at different time instances. 
This considers a case with $0.35\,\mbox{nano grams}$ of surfactant molecules deposited on a planar film. 
A finite difference solution of (\ref{eq:cd}) is depicted by the solid line, whereas, symbols denote data from MD.
The close agreement affirms the validity of the continuum model in capturing the nanoscale transport mechanism. 
The dotted lines represent the spline fit, $\zeta$ to the instantaneous liquid-vapor surface.  
Section $3.4$ of the \textit{Supplementary Material} includes similar plots for other amounts of deposited surfactants.
It is observed that simulations with denser monolayer, align more closely with the numerical solution of the transport equation. 
The reasons for this are not obvious, but could be due in part to uncertainties associated with estimating the diffusion coefficient in these surfactant-deficient systems.
Also, for less denser monolayers, local concentration fluctuations are likely, which would not be captured by the mean-field continuum model.

\begin{figure}[!t]
    \centering
    \includegraphics[width=0.98\linewidth]{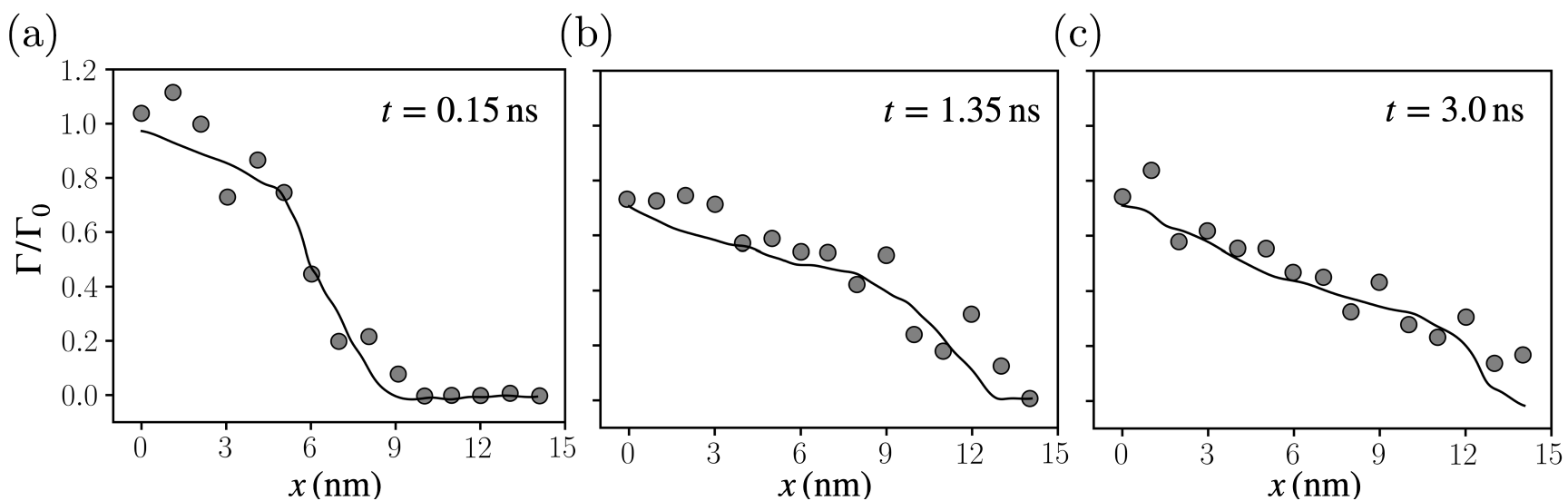}
    \caption{Temporal variation of surfactant concentration on a droplet. Data from molecular dynamics (MD) simulations are represented by symbols, while the solid lines indicate the numerical solutions to the transport equation.}
    \label{fig:droplet}
\end{figure}

In addition to the spreading of surfactant across a planar film, the analysis has been extended to a globally curved geometry, i.e., a droplet as in Fig. \ref{fig:shcematic} (b). 
Although a droplet has a higher global curvature than the planar films, Fig. \ref{fig:droplet} shows that the transport mechanism remains largely unaffected.
For planar films, the far-right end of the film remained devoid of surfactant, resulting in a persistent surface tension gradient that favored transport throughout the simulation. In contrast, for the droplet case, surfactants from both the upper and lower regions of the droplet redistributed, reducing the area of surfactant-poor regions. Consequently, the transport process quickly approached a state where surfactants from both halves of the droplet came into close proximity, significantly diminishing the driving force for further transport and nearly halting any directional flow of the surfactants.

The diffusion coefficients of the surfactant molecules are higher for low initial concentrations. 
At higher concentrations, the relative impact of self diffusion diminishes, potentially due to the collective motion of the monolayer resulting from the increased inter-surfactant interactions. This suggests, in a dilute system, the diffusion contribution, $\eta_\mathrm{diff.} (t)$ might have a relatively greater impact.
The contributions from the geometric term, $\eta_\mathrm{geom.} (t)$ is consistently insignificant across all cases,
possibly because both $\kappa_i$ and $v_i$ fluctuate around zero, and are small numbers, rendering their product an order of magnitude smaller than the advective and the diffusive terms. They might gain more importance for surfaces with higher local curvatures or when the surface is on a growing/shrinking droplet, film or bubble. These contribution terms, plotted in Fig. \ref{fig:relative}(a), are defined in the \textit{Supplementary Material}.

\subsection{Spreading length}
\begin{figure}[!t]
    \centering
    \includegraphics[width=0.98\linewidth]{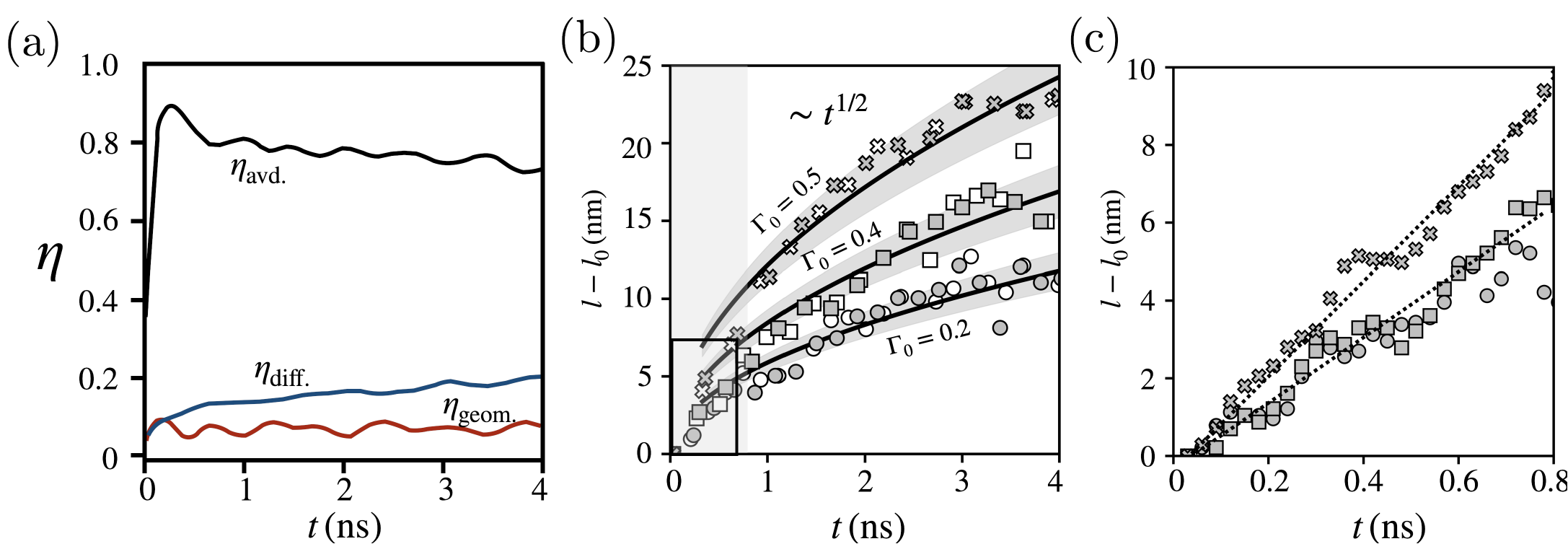}
    \caption{(a) Relative contributions of advection, diffusion and geometrically induced transport.
    (b) Spreading of monolayer: symbols denote MD data, with empty and solid symbols denoting independent simulations. Solid lines show $l\sim t^{1/2}$ fitting to the data. Shaded region shows $90\%$ confidence interval of the fitting. (c) Zoomed in view of the initial rapid transient period (boxed in panel b) showing the early time (linear) spreading.}
    \label{fig:relative}
\end{figure}
A more pragmatic parameter to characterize surfactant transport is the spreading rate which essentially denotes the rate at which the front of the surfactant layer spreads across the liquid-vapor interface. 
The presence of a monolayer creates a difference in surface tension across the interface of the underlying liquid layer. 
This tension gradient drives the spreading of the monolayer. The spreading must overcome the viscous drag, and by balancing these two forces, \citet{ahmad1972simple} presented a simplified scaling law for the spreading length, $l$, which has the form, $l\sim t^{1/2}$. 
Despite such simplicity, the authors observed good agreement with experimental data. 
Fig. \ref{fig:relative} (b) shows that the spreading of the surfactant monolayer follows the characteristic square root dependency on time for a range of initial concentrations. 
This observation agrees with past experiments studies, for example, \citet{hanyak2012soluble} studied the spreading of SDS on deformable liquid surfaces, and obtained an average exponent of $0.48$ for various surfactant concentrations. 
Similar power law dependency was observed in several other studies \citep{borgas1988monolayer,Hamraoui2004,hernandez2015marangoni,mitra2016understanding}.  
The scaling exponent may assume a range of values depending on the geometry of the problem, and the physicochemical properties of the liquids involved, such as viscosity, surface tension, miscibility, and surface contamination \citep{rafai2005spreading, santiago2001spreading,Lee2007}.

While the $t^{1/2}$ rate dependence persists over the whole simulation time window in the present study, the initial spreading ($t<0.5\, \mathrm{ns}$) appears to be relatively faster with the exponent $n\to1$, see panel (c) for a zoomed view of the initial spreading period. 
Similar accelerated initial spreading was also observed in several previous experimental studies of monolayer spreading \citep{Lee2007,mitra2016understanding,motaghian2022surfactant}.
This local ballistic or molecular-scale relaxation behavior may be a unique feature of the molecular description provided by MD \citep{alder1967velocity}, although its origin is still not entirely clear and would be worthy of further investigation. It is worth mentioning that a ballistic to diffusive crossover is expected as interactions and scattering of the particles attain greater importance \citep{Huang2011}.

\vspace{-2mm}
\section{Conclusions}
This study uses empirical data from NEMD simulations to parameterize a continuum model of surfactant transport, confirming its efficacy at the molecular scale. The advection-diffusion equation, a cornerstone of fluid mechanics, is affirmed to be an effective descriptor of the complex interfacial behaviors studied here. 
The findings highlight the importance of considering the local and temporal variations of the surface fluxes, surface deformation, and the local gradients of surfactant concentration in modeling the transport of surfactant monolayers.  The agreement between the outcomes of our NEMD simulations and the continuum model provides strong validation for the applicability and robustness of the continuum framework for nanoscale phenomena—a domain where such methods often reveal their limitations.
In addition, a novel and exact molecular-scale transport equation is derived. This work thus bridges the gap between molecular dynamics and continuum theories, offering a comprehensive understanding of surfactant transport across scales.

\bibliographystyle{unsrtnat}
\bibliography{references}

\section{Acknowledgments}
MRR was supported by Shell, and the Beit Fellowship for Scientific Research. JPE was supported by the Royal Academy of Engineering (RAEng). 
LS thanks EPSRC for a Postdoctoral Fellowship (EP/V005073/1).
DD acknowledges a Shell/RAEng Research Chair in Complex Engineering Interfaces and the EPSRC Established Career Fellowship (EP/N025954/1).
Authors are grateful to the UK Materials and Molecular Modelling Hub for computational resources funded by EPSRC (EP/T022213/1, EP/W032260/1 and EP/P020194/1). 
Authors thank Erik Weiand for help in assigning the MARTINI parameters, and Carlos Corral-Casas, Carlos Braga and Stavros Ntioudis for fruitful discussions.

\vspace{-2mm}
\section*{Author Contributions}
\noindent DD, ERS, and MRR conceptualized the problem, DD and LS acquired funding, MRR developed the methodology with contributions from JPE and ERS. MRR carried out the MD and FD simulations, and analyzed the results. ERS derived the molecular-scale transport equation. MRR drafted the manuscript with contributions from ERS. DD, ERS, JPE and DMH supervised the project. All authors discussed the results, and edited the manuscript.


\end{document}


\title{Supplementary Material for  Nanoscale Surfactant Transport: Bridging Molecular and Continuum Models}

\author{Muhammad Rizwanur Rahman${^{1,*}}$, James P. Ewen ${^1}$, Li Shen${^1}$, D. M. Heyes${^1}$, Daniele Dini${^1}$, and E. R. Smith${^2}$}
\affil{${^1}$Department of Mechanical Engineering, Imperial College London, South Kensington Campus, London SW7 2AZ, United Kingdom \newline ${^2}$Department of Mechanical and Aerospace Engineering, Brunel University London, Uxbridge, London UB8 3PH, United Kingdom \newline ${^*}$ Email: m.rahman20@imperial.ac.uk}

\maketitle

\tableofcontents
\section{Molecular Dynamics Simulations}
\begin{figure} [!t]
    \centering
    \includegraphics[width=0.98\linewidth]{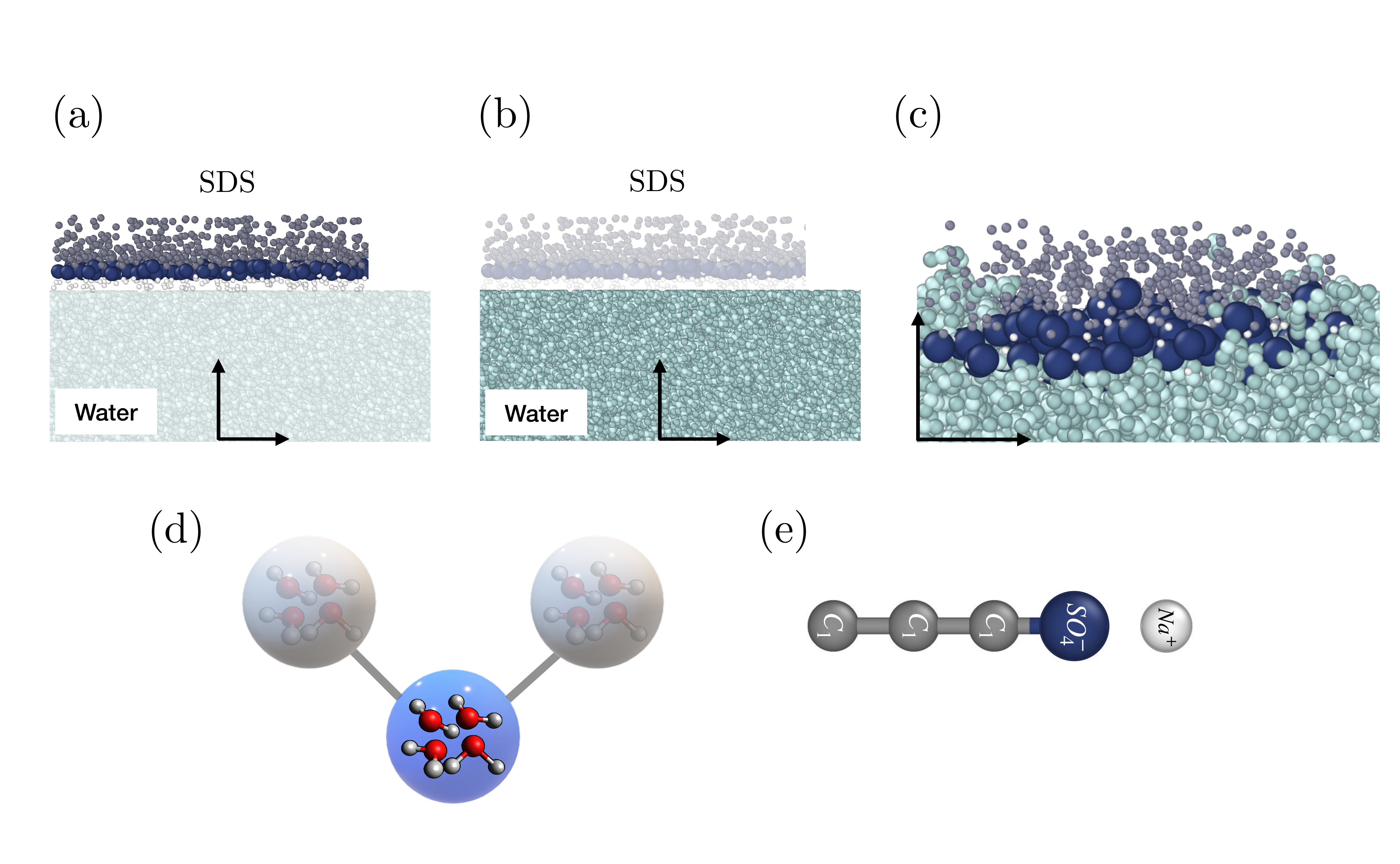}
    \caption{(a,b) Initial configuration of SDS layer and water film, with Hydrophilic head group of SDS close to the water surface, and hydrophilic tail group away from the surface. These were separately equilibrated: equilibration of (a) SDS, (b) water. (c) Zoomed view of a portion of the surfactant-water interface after equilibration. 
    Schematic of the coarse grain description of (d) MARTINI polarizable water model consisting of three beads, where 4 water molecules are mapped into one bead, and (e) SDS molecule.  }
    \label{fig:initialconfig}
\end{figure}

The MARTINI water model is composed of three particles: a charge-neutral central particle (\(w\)), accompanied by positively (\(wp\)) and negatively charged (\(wm\)) particles. The central particle, \(w\) engages in Lennard-Jones interactions with surrounding particles, whereas \(wp\) and \(wm\) interact via Coulombic forces, with no interaction occurring within the same water bead. The mass of each coarse-grained bead is set at 72 amu, reflecting the aggregate mass of four actual water molecules, see Fig. S\ref{fig:initialconfig}.
The initial configurations were prepared using the open source codes Packmol \citep{martinez2009packmol} and Moltemplate \citep{jewett2021moltemplate}.
All simulations for this study were carried out in the Large-scale Atomic/Molecular Massively Parallel Simulator, LAMMPS \citep{thompson2022lammps}. The molecular model incorporated detailed representations for both bonded and non-bonded interactions. Bonded interactions were described using harmonic potentials for bonds, i.e., 
\begin{displaymath}
    V_b (r_{ij}) = \frac{1}{2} k_{ij}^b (r_{ij} - b_{ij})^2 
\end{displaymath}
were, $V_b (r_{ij})$ is the potential energy associated with the bond between atoms $i$ and $j$ at distance $r_{ij}$. $b_{ij}$ is the equilibrium bond length between atoms $i$ and $j$, i.e., the natural length of the bond at which the potential energy is minimized, and $k_{ij}$ is the force constant that determines the stiffness of the bond. 

To model the energy associated with the bending of angles between three consecutive bonded atoms, the cosine-squared potential was used. This is generally expressed as: 
\begin{displaymath}
    V_a (\theta) = \frac{1}{2} k^\theta \left( \mbox{cos}\theta  - \mbox{cos}\theta_0 \right)^2
\end{displaymath}
here, $V_a$ is the potential energy as a function of angle 
$\theta$ formed by three bonded atoms $i,j\,\mathrm{and}\, k$; $\theta_0$ is the equilibrium angle, and $k^\theta$ is the force constant for angle. 
Non-bonded interactions were accounted for through a combination of Lennard-Jones and Coulombic potentials, employing a hybrid overlay approach to capture the nuanced interplay of forces at different scales. Generally expresses as: 
\begin{displaymath}
    V_\mathrm{non-bonded} (r) = V_\mathrm{LJ} (r) + V_\mathrm{Coul.} (r)
\end{displaymath}
where $ V_\mathrm{LJ} (r)$ is the Lennard-Jones potential, i.e., 
\begin{displaymath}
    V_\mathrm{LJ} (r) = 4\epsilon \left[ \left(\frac{\sigma}{r}\right)^{12} - \left(\frac{\sigma}{r}\right)^6 \right]
\end{displaymath}
and the Coulombic interactions are given by, 
\begin{displaymath}
   V_\mathrm{Coul.} (r) = (q_i q_j/ 4\pi\epsilon_0 \epsilon_r) \frac{1}{r} 
\end{displaymath}
where, symbols carry their usual meanings. 
The long-range electrostatic forces were calculated by  \textit{Particle-Particle Particle-Mesh} (PPPM) algorithm. This method allowed us to accurately account for the electrostatic component of our system at reasonable computational cost.  \textit{SHAKE} algorithm was used to maintain rigid $O-H$ bonds within water molecules. \newline

The initial configurations were separately equilibrated (Fig. S\ref{fig:initialconfig}\,panels a$-$c) in the NVT (canonical) ensemble to stabilize the system to a target temperature of 300 K.   The equilibration of the two species was carried out separately to restrict any surfactant transport before the system is stabilized, and before the production phase starts.
Following this, the system underwent a NVT production phase when data were collected for analysis. Panels (d,e) show, respectively, the coarse grain descriptions of water and SDS.

\section{Derivation of a Molecular Equivalent of the Transport Equation}
In what follows, we aim to derive a molecular version of the  surfactant transport equation of \citet{scriven1960marangoni},
\begin{align}
\frac{d\Gamma}{dt}   + \boldsymbol{\nabla}_s \cdot {\Gamma \boldsymbol{u}_s} + \frac{\Gamma \dot{a}}{2a} = j_e
\label{Shriven}
\end{align}
Equation\,(\ref{Shriven}) describes the time evolving surface advection, surfactant exchange with the bulk, and also accounts for any surface expansion in time.
Taking the expression for a surfactant concentration at any point in the domain,
\begin{align}
\Gamma(x,y,z,t) =  \displaystyle\sum_{i=1}^{N_S} m_i \delta(\boldsymbol{r}- \boldsymbol{r}_i) 
\label{Gamma}
\end{align}
where the Dirac delta functions includes the molecules when their positions $\boldsymbol{r_i} = [x_i, y_i, z_i]$ are located at $\boldsymbol{r} = [x, y, z]$.
In practice, an average of time or an ensemble of systems would then be required.
In a molecular system, all quantities must also be averaged over a region in space, so the control volume is the most natural form to work in.
The integral over a 3D control volume, which follows an interface $\xi(x,y,z,t)$ with the corners $x^{\pm}$, $y^{\pm}$ and $z^{\pm}$ being functions of $\xi$, is then, 
\begin{align}
\int_V \Gamma(x,y,z,t) \,dV & =  \int_V \displaystyle\sum_{i=1}^{N_S} m_i \delta(\boldsymbol{r}- \boldsymbol{r}_i) dV 
\nonumber  \\
& = \int_{x^-}^{x^+} \int_{y^-}^{y^+} \int_{z^-}^{z^+}  \displaystyle\sum_{i=1}^{N_S} m_i \delta(x - x_i) \delta(y - y_i) \delta(z - z_i) \, dz \, dy \, dx 
\nonumber  \\
& =
\int_{\chi^-}^{\chi^+} \int_{\psi^-}^{\psi^+} \int_{\omega^-}^{\omega^+} \displaystyle\sum_{i=1}^{N_S} m_i \delta(\chi - \chi_i) \delta(\psi - \psi_i) \delta(\omega - \omega_i)  J \, d\chi \, d\psi \, d\omega \nonumber \\
& = \displaystyle\sum_{i=1}^{N_S} m_i  \vartheta_i J_i
\end{align}
where we have used the property that the Dirac delta function at any point in vector space $\delta(\boldsymbol{r}- \boldsymbol{r}_i)$ is the product of the three orthogonal vectors, either the three Cartesian coordinates or equivalently it could be written in terms of coordinates aligned with the two tangential direction to surface $\xi$ at every point, denoted as $\chi$ and $\psi$ and the surface normal $\omega$. 
The intergral is then changed using a Jacobian mapping $J=J(\chi, \psi, \omega,t)$, to a constant cuboidal control volume of length $\Delta \chi$, width $\Delta \psi$ and height $\Delta \omega $.
This is centered on the interface $\xi$ with half its total volume $\Delta \omega/2$ either side, and length/width constant values $  \Delta \chi/2$ and $\Delta \psi/2$ respectively. The notation for the cuboid limits denote $\chi^{\pm} = \chi \pm \Delta \chi/2$ and similar for $\psi$ and $\omega$.

The sifting property of the Dirac delta function $\int \delta (x-a) f(x) = f(a)$ results in a Jacobian $J_i = J(\chi_i, \psi_i, \omega_i,t)$ evauated at the location of each molecule in the sum.
The integral between constant limits is called the control volume function $\vartheta_i$, the product of three boxcar functions with $\vartheta_i \equiv \Lambda_{\chi} \Lambda_{\psi} \Lambda_{\omega}$ each the difference of two Heaviside functions H, $\Lambda_{\chi} \equiv (H(\chi^+ - \chi_i)-H(\chi^- - \chi_i))$ and similar for $\psi$ and $\omega$ so the product is a cuboid in 3D $(\chi,\psi,\omega)$ space.
This has the property that the derivative of $\vartheta_i$ in a given direction, is a function which selects molecules crossing a square face of the cuboid in $[\chi,\psi,\omega]$ space, that is $\partial \vartheta_i / \partial \chi =  \left[\delta(\chi^+ - \chi_i)-\delta(\chi^- - \chi_i)\right]\Lambda_{\psi} \Lambda_{\omega}$. 
When mapped back to real space, this is then a crossing of a molecule on the control volume face as it moves with the interface, for $\chi$ and $\psi$ along the interface and for $\omega$ a flux over the interface. \newline 

\begin{figure} [!t]
\centering
\includegraphics[width=0.8\textwidth]{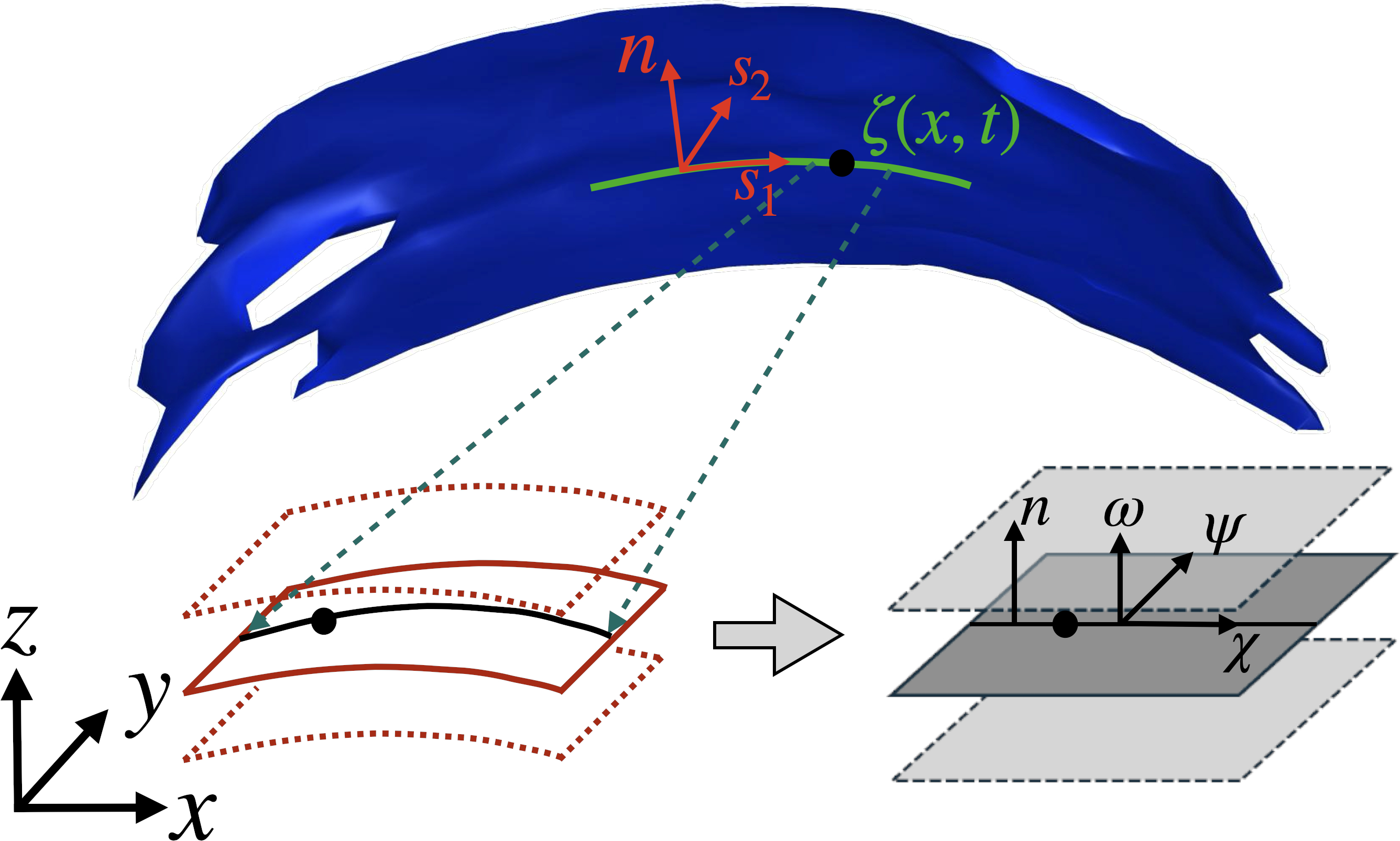}
\caption{
Schematic diagram showing mapping of a deforming surface from $(x, y, z)$ space to $(\xi, \psi, \omega)$ space.
}
\label{mapping_plots}
\end{figure}

To express the conservation of surfactants in terms of these fluxes, the control volume function can be given the same treatment as a cuboidal control volume \citep{smith2012control}, linking time evolution in a volume to molecular surface flux terms.
The Jacobian has all the time dependence of the moving volume $J_i =J(\chi_i, \psi_i, \omega_i, t)$ and only the molecular positions are functions of time, i.e. $d H\left[\chi - \chi_i(t)\right]/dt = -\dot{ \chi}_i \delta \left[\chi - \chi_i(t)\right]$.
The time evolution of a control volume on the interface is then as follows,
 \begin{align}
\frac{d}{dt}\int_V \Gamma(x,y,z,t) \,dV  &=
\frac{d}{dt}\displaystyle\sum_{i=1}^{N_S} m_i J_i \vartheta_i 
\nonumber \\
& =\displaystyle\sum_{i=1}^{N_S}m_i \left[ J_i\dot{\Lambda}_{\chi} \Lambda_{\psi} \Lambda_{\omega}  + J_i\Lambda_{\chi} \dot{\Lambda}_{\psi} \Lambda_{\omega}  +  J_i\Lambda_{\chi} \Lambda_{\psi} \dot{\Lambda}_{\omega}  + \dot{J}_i\Lambda_{\chi} \Lambda_{\psi} \Lambda_{\omega} \right]
\nonumber \\
& = \displaystyle\sum_{i=1}^{N_S}m_i \left[ - J_i \dot{\chi}_i dS_{\chi i} - J_i \dot{\psi}_i dS_{\psi i} - J_i \dot{\omega}_i dS_{\omega i} + \dot{J}_i \vartheta_i  \right]
\end{align}
where we introduce the notation $dS_{\chi i} \equiv dS_{\chi i}^+ - dS_{\chi i}^-= \left[\delta(\chi^+ - \chi_i)-\delta(\chi^- - \chi_i)\right]\Lambda_{\psi} \Lambda_{\omega}$, the flux of molecules in the $\chi_i$ direction crossing surfaces at $\chi^{\pm}$. In mapped space, this calculation is simply checking if mapped molecule position $\chi_i$ crosses a flat surface $\chi^+$ or $\chi^-$ as shown in Figure S\ref{mapping_plots} a), but in real space this represents flow along the tangent to the surface $\xi$, as shown in Figure S\ref{mapping_plots} b) where the end surfaces are normal to the surface.
Collecting all terms on one side of the equation and dividing by $J_i$,
\begin{align}
 \displaystyle\sum_{i=1}^{N_S} m_i  \left[\frac{1}{J_i} \frac{d}{dt} (\vartheta_i J_i) + \dot{\chi}_i dS_{\chi i} + \dot{\psi}_i dS_{\psi i} + \dot{\omega}_i dS_{\omega i} - \vartheta_i \frac{\dot{J}_i}{J_i} \right] = 0
\label{time_eqn}
\end{align}
allow us to compare terms to the continuum form of \eqref{Shriven}.
Note, we have used the control volume form as this is the most sensible for a molecular system, because molecules need a finite volume in space to average over.
However, in order to understand the link to the differential form of \eqref{Shriven}, we can take the limit that this volume tends to zero. Given the small size of the molecular volumes this is reasonable from a continuum perspective.
In the limit the volume length $\Delta \chi$ tends to a zero in the $\chi$ direction, the limits approach each other so the surface flux term $ \lim_{\Delta \chi \to 0} dS_{\chi i} = \frac{\partial }{\partial \chi} \delta(\chi - \chi_i) \Lambda_{\psi}$ becomes the derivative of a delta function \citep{smith2012control}. 
The similar limit of a boxcar function is simply the delta function, $\lim_{\Delta \psi \to 0}\Lambda_{\psi} = \delta (\psi - \psi_i)$.
In this form, the clear link to advection term in the continuum equations is apparent, as it represents the derivative in the tangential direction of the velocity along the tangent time mass of surfactant,
 \begin{align}
\lim_{\substack{\Delta \chi \to 0 \\ \Delta \psi \to 0 \\ \Delta \omega \to 0}}\displaystyle\sum_{i=1}^{N_S}m_i  \left [\dot{\chi}_i dS_{\chi i} + \dot{\psi}_i dS_{\psi i} \right]
& = \displaystyle\sum_{i=1}^{N_S}m_i  \left[\frac{\partial }{\partial \chi}  \dot{\chi}_i + \frac{\partial }{\partial \psi} \dot{\psi}_i \right] \delta(\chi - \chi_i) \delta(\psi - \psi_i)\delta(\omega - \omega_i)  
\nonumber \\ 
&= \frac{\partial }{\partial \boldsymbol{s}} \cdot  \displaystyle\sum_{i=1}^{N_S}m_i   \dot{\boldsymbol{s}}_i  \delta(\tilde{\boldsymbol{r}} - \tilde{\boldsymbol{r}}_i) = \boldsymbol{\nabla}_s \cdot {\Gamma \boldsymbol{u}_s}
 \label{advection}
\end{align}
where $\boldsymbol{s} = [\chi, \psi]$, and $\tilde{\boldsymbol{r}} = [\chi, \psi, \omega]$. We emphasis that the delta function is always defined on the interface.
Equation (\ref{advection}) is analogous to \citet{irving1950statistical} who defined gradient of momentum as $\nabla \cdot \rho \boldsymbol{u} = \frac{d}{d\boldsymbol{r}}  \sum_{i=1}^{N} \langle m_i \dot{\boldsymbol{r}}_i \delta(\boldsymbol{r} - \boldsymbol{r}_i) ;f \rangle$. \newline 

For the $\dot{\omega}_i dS_{\omega i}$  term in (\ref{time_eqn}), the limit of zero volume would give a derivative in the surface normal direction,
 \begin{align}
 \lim_{\substack{\Delta \chi \to 0 \\ \Delta \psi \to 0 \\ \Delta \omega \to 0}}\displaystyle\sum_{i=1}^{N_S}m_i  \dot{\omega}_i dS_{\omega i} = 
\frac{\partial }{\partial \omega}\displaystyle\sum_{i=1}^{N_S}  m_i \dot{\omega}_i  \delta(\tilde{\boldsymbol{r}} - \tilde{\boldsymbol{r}}_i) = j_e
 \end{align}
To understand the link to the surfactant source/sink term $j_e$, it is clearer to interpret this in terms of fluxes. The surfactant flux over $\omega^+$ given by $m_i \dot{\omega}_i dS_{\omega i}^+$ can be assumed to be zero, as this surface is above the interface which would be evaporation of surfactant. The flow of surfactant over the bottom surface, $m_i  \dot{\omega}_i dS_{\omega i}^- $, is therefore the surfactant movement from bulk to the surface, a source/sink term in the surface equations.
And the final term, the deformation of the surface in time, denoted as geometric effect in the main manuscript, is given by the time derivative of the Jacobian $\dot{J}$,
\begin{align}
\lim_{\substack{\Delta \chi \to 0 \\ \Delta \psi \to 0 \\ \Delta \omega \to 0}}\displaystyle\sum_{i=1}^{N_S}m_i  \vartheta_i  \frac{\dot{J_i}}{J_i} =\displaystyle\sum_{i=1}^{N_S} m_i \delta(\tilde{\boldsymbol{r}} - \tilde{\boldsymbol{r}}_i) \frac{\dot{J}_i}{J_i} = \Gamma\frac{ \dot{a}}{2a} 
\end{align}
which is simply the definition of $\Gamma$ from equation \ref{Gamma} multiplied by the Jacobian at the position of each molecule.
This means, in the molecular version of \citet{scriven1960marangoni}, the total deformation of the surface is the sum of the deformation at the location of each molecule.
The equivalent of each term in the molecular equation is summarized here,
\begin{equation}\label{eq:molecular_SI}
    \sum_{i=1}^{N_S}  \bigg[ \underbrace{\frac{1}{J_i} \frac{d}{dt} \left( m_i\vartheta_i J_i \right)}_{\mathrm{unsteady}} 
    + \underbrace{ m_i\dot{\chi}_i dS_{\chi i} + m_i\dot{\psi}_i dS_{\psi i} \vphantom{\frac{1}{J_i}}}_{\mathrm{advection}} \;\; - \!\!\!\!\underbrace{m_i\vartheta_i  \frac{\dot{J_i}}{J_i}}_{\mathrm{geometric\, effect}} \!\!\!\!\!\! \bigg] =  \sum_{i=1}^{N_S} \underbrace{m_i\dot{w_i} dS_{wi}^-\vphantom{\frac{1}{J_i}}}_{\mathrm{source\, term}} 
\end{equation} \newline 
Or in the limit of zero volume, we get the equivalent form to the continuum differential equation,
\begin{equation}
     \underbrace{\sum_{i=1}^{N_S} \frac{1}{J_i} \frac{d}{dt} \bigg[ m_i J_i \delta(\tilde{\boldsymbol{r}} - \tilde{\boldsymbol{r}}_i)   \bigg]}_{\mathrm{unsteady}} 
    + \underbrace{ \frac{\partial }{\partial \boldsymbol{s}} \cdot  \sum_{i=1}^{N_S} m_i   \dot{\boldsymbol{s}}_i  \delta(\tilde{\boldsymbol{r}} - \tilde{\boldsymbol{r}}_i) \vphantom{\frac{1}{J_i}}}_{\mathrm{advection}}  - \underbrace{ \sum_{i=1}^{N_S} m_i  \frac{\dot{J}_i}{J_i} \delta(\tilde{\boldsymbol{r}} - \tilde{\boldsymbol{r}}_i) }_{\mathrm{geometric\, effect}} =   \underbrace{je \vphantom{\sum_{i=1}^{N_S}}}_{\mathrm{source\, term}}  \nonumber
\end{equation}

The surface flux in equation (\ref{eq:molecular_SI}) can be obtained directly by counting molecular crossings, which is the velocity method of planes \citep{todd1995pressure}.
However, the statistics are known to be much worse for surface flux measurements when compared to volume averaging, especially for the thin volumes used on an interface. Instead, it is also possible to take averages in the volume to give better statistics, knowing the surface fluxes are equivalent to the 
volume measurements in the limit that the volumes are small \citep{heyes2011equivalence}.
For example, if the MD surfactant concentration in a cell at time step, $t$ is:
\begin{equation}
[\Gamma u_s ]_{I+1}^{t} = \sum_{i=1}^{N_s} m_i \dot{s}_i \vartheta_i[s+\Delta s, t]  
\end{equation}
where we shift the position of the control volume function, $\vartheta_i[s+\Delta  s,t+1] = [H(s^+ + \Delta s – s_i(t)) - H(s^- + \Delta s – s_i(t))]\Lambda_{\omega}$, noting we have used scalar $s$ in place of $\chi$ in this 1D example. 
Then the MD advection term can be approximated as a finite difference style algorithm,
\begin{equation}
\frac{\partial }{\partial \boldsymbol{s}} \cdot  \displaystyle\sum_{i=1}^{N_S}m_i   \dot{\boldsymbol{s}}_i  \delta(\tilde{\boldsymbol{r}} - \tilde{\boldsymbol{r}}_i) \approx \frac{\left([\Gamma u_s ]_{I+1}^{t} - [\Gamma u_s]_{I}^{t}\right)}{\Delta s}  
\end{equation}
Similarly, it is possible to define the unsteady term in the same way.

\section{Numerical Method}
This section describes the details of how the continuum surfactant transport equation is discretized using FD. 
We discretise the evolution of $\Gamma$ in time using a simple backwards Euler,
\begin{equation}\label{eq:Numerics_dt}
\frac{\partial \Gamma}{\partial t} \approx \frac{\Gamma_{I}^{n} - \Gamma_{I}^{n-1}}{\Delta t}
\end{equation}
where superscripts denote time steps, and subscripts denote spatial cells. 
The advection terms is one dimensional in $s$, hence, $\nabla_s \cdot (\Gamma \mathbf{u_s}) = \partial \Gamma u / \partial s$ with the subscript in $u_s$ dropped for notational simplicity,
\begin{equation}\label{eq:Numerics_adv}
\frac{\partial \Gamma u }{\partial s}= u \frac{\partial \Gamma  }{\partial s} + \Gamma \frac{\partial u  }{\partial s} \nonumber \\
\approx  u_I \frac{\Gamma_{I+1} - \Gamma_{I}  }{\Delta s} + \Gamma_I \frac{u_{I+1} - u_{I}}{\Delta s} 
\end{equation}
Here, forward difference has been used to include upwinding \citep{hirsch2007numerical} as $u>0$ with surfactant moving from left to right, and $\Gamma \ge 0$. The sum of this term over all $N_\text{cells}$ gives the total advection $\eta_\text{adv.}$ at each time plotted in the main text.
For the diffusion term, a second order discretisation is applied,
\begin{equation}\label{eq:Numerics_diff}
D_s \frac{\partial^2 \Gamma}{\partial s^2}  \approx D_s \frac{\Gamma_{I+1} - 2\Gamma_{I} + \Gamma_{I-1}}{(\Delta s)^2}
\end{equation}
The sum of $I$ over all $N_\text{cells}$ gives the total diffusion $\eta_\text{diff.}$ at each time.
Together, the discretized form of the transport equation is:
\begin{equation}\label{eq:Numerics_full}
\frac{\Gamma_{I}^{n} - \Gamma_{I}^{n-1}}{\Delta t} + u_I^n \frac{\Gamma_{I+1}^n - \Gamma_{I}^n}{\Delta s} + \Gamma_I^n \frac{u_{I+1}^n - u_{I}^n}{\Delta s}  + \Gamma_{I}^n \kappa_I^n v_I^n = D_s \frac{\Gamma_{I+1}^n - 2\Gamma_{I}^n + \Gamma_{I-1}^n}{(\Delta s)^2} 
\end{equation}
Assuming a zero source term (i.e., $j_n = 0$), the discretized form of the transport equation,  with like terms grouped together in the form of a matrix $a \Gamma = b$, is given by
\begin{align}
    \Gamma_{I+1}^n \overbrace{\left[ \frac{\Delta t u_I^n}{\Delta S}  - \frac{D \Delta t}{(\Delta S)^2}  \right]}^\mathrm{a_{I+1}^n} + 
    \Gamma_I \overbrace{\left[1  + \Delta t \left( \frac{u_{I+1}^n - 2u_I^n}{\Delta s} + \kappa_I^nv_I^n + \frac{2D}{(\Delta s)^2}  \right) \right]}^{\mathrm{a_I^n}} \nonumber \\
    + \Gamma_{I-1}^n \overbrace{\left[ -\frac{D \Delta t}{(\Delta S)^2} \right]}^\mathrm{a_{I-1}} = \Gamma_I^{n-1}
\end{align}
here, $I$ and $n$ denotes, respectively, spatial and temporal discretization. 
The system of equations is solved using the inbuilt linear algebra solver in Numpy (v.1.26.0). \newline

The accuracy of the implicit numerical scheme has been confirmed by testing against the analytical solutions of pure convection and pure diffusion. 
In solving the surfactant transport, the domain assumes symmetry about $x=0$, and hence is discretized from $x=0$ to $x=l$ along the positive $x$-axis. The initial concentration field was set to the MD data at $t=t_0$, i.e., 
$\Gamma_{0,FD} =\Gamma_{0,MD}$. 
The left boundary condition was modeled as
$\Gamma (x=0,t) = \alpha e^{-\lambda t} + \alpha_0 $ based on a fit to the MD data. At the far right, $\Gamma (x=l,t) =0$ as surfactant molecules do not reach the right end of the film during the simulation time.\newline

The local diffusion coefficient $D (s,t) = 3.2\,e^{-4N/3}$ for finite $N$, was obtained by measuring the diffusion coefficient as a function of surfactant (number) concentration (see SI section \ref{sec:diffusion}). 
The effective diffusion coefficient, which dynamically updates based on the local concentration, reflects non-linear interactions and is crucial for accurately modeling transport in systems with variable properties.
Note, the diffusion coefficients obtained in our simulations are larger than the experimentally measured bulk diffusion of SDS in aqueous solutions \citep{ribeiro2003diffusion} by a factor of $2$ to $3$. This can be attributed to the speed up process of the kinetics in coarse grain models which contributes to the over-estimation of diffusion \citep{vogele2015properties, schmitt2023comparison}. In Martini models, the best estimated kinetic speed up factor is approximately $4$, however, the polarized water model is reported to show approximately $2.5$ times speed up \citep{marrink2013perspective}. However, the present study does not aim to predict these values, rather the focus is to capture the overall transport mechanism. 
To overcome any potential mass imbalance arising from numerical diffusion or discretization errors, a conservation correction mechanism is integrated to ensure that the total quantity of the transported scalar is preserved throughout the simulation, i.e., $\int_0^l \Gamma ds = \mathrm{Constant}$. \newline 

The numerical resolution of the velocity field along the local tangential direction of a curved surface was a crucial aspect of the modeling process. 
This involves computing the scalar component of the velocity that is tangential to the topology of the surface at any given point. 
This has been done by resolving the local velocity into the normal and tangential components relative to the surface.  
To get the local curvature of the surface for the geometric term, the deforming liquid-vapor surface is identified from the MD based on a density threshold. The surface is then fitted to a spline using the \textit{UnivariateSpline} function from \textit{scipy} library to obtain a functional $\xi = \xi(x,z,t)$.
Using the derivatives of the spline, the distance $ds$ as well as the normal and the tangential vectors are obtained at each time step. 
These derivatives of the surface at each point are used in,
\begin{align} 
\kappa= \frac{\partial^2 \xi}{ \partial x^2}  \left(1 + \left[\frac{\partial \xi}{ \partial x} \right]^2 \right)^{-\frac{3}{2}} 
\end{align}
which is the sum of the principal curvatures.
The FD expression for calculating the geometetric term is then as follows:
\begin{equation}
    \eta_\text{geom.} = \displaystyle\sum_{I=1}^{N_\text{cells}} \left([\kappa \Gamma v]^t_{I+1} - [\kappa \Gamma v]^t_I]\right)/\Delta s
\end{equation}

\section{Additional Results and Discussions}
\subsection{Surface deformation} 
\begin{figure}  
    \centering
    \includegraphics[width=1\linewidth]{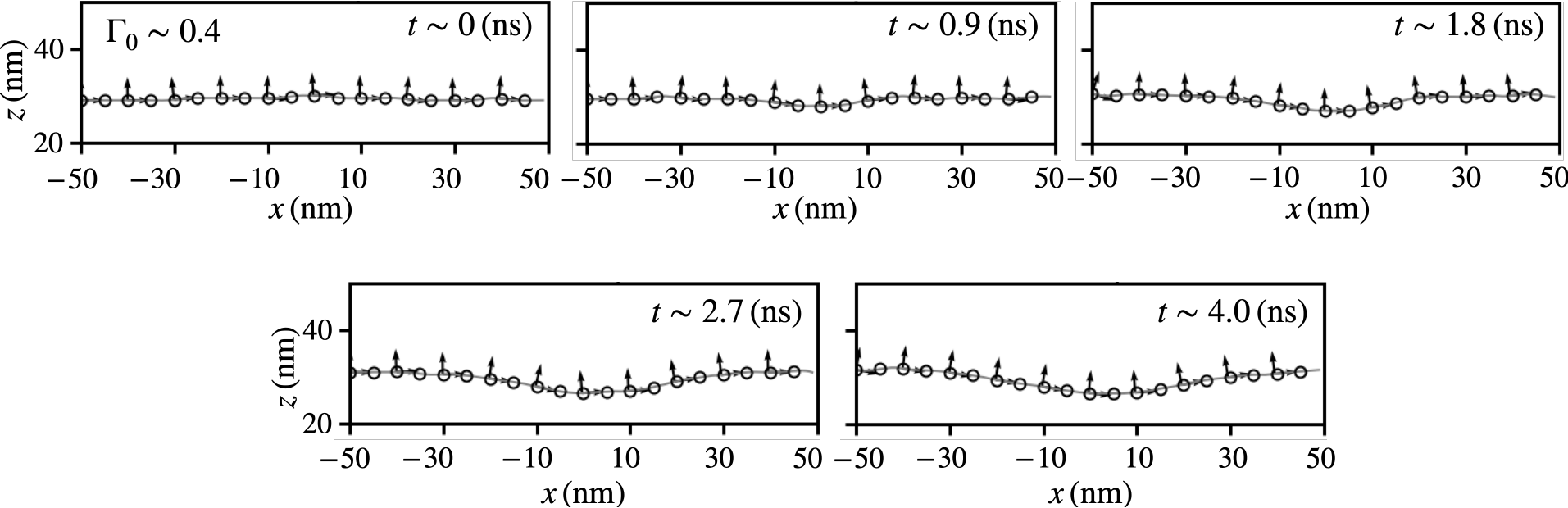}
    \caption{Evolution of the surface of a film with a monolayer of initial concentration, $\Gamma_0 \sim 0.4$, placed at the central area ($x=-10\,\,\mathrm{to}\,\,10$) of the surface. Symbols denote surface position from MD data, while faint gray lines denote fitting to the data. Arrows show local directions normal and tangential to the surface.}
    \label{fig:surfacedeformation}
\end{figure}

Figure S\ref{fig:surfacedeformation} shows the time evolution of surface deformation in a thin liquid film induced by the lateral migration of surfactant molecules. 
The top surface of the water layer was fitted to a cubic spline which ensures ensures continuity up to the second derivatives.
At the initial stage ($t\approx 0$ ns), the surface presents a near-uniform profile. As the surfactant molecules spread across the surface, they locally decrease the surface tension in the central region, giving rise to Marangoni flow. 
This dynamics result in the advective transport of water molecules away from the central zone. Consequently, as in Fig. S\ref{fig:surfacedeformation}, a notable reduction in film thickness at the central zone becomes evident with simultaneous increase at the edges of the film. Arrows denote local surface normal, and tangent to the surface. Surfactant velocities were resolved along these directions for the numerical solution of the transport equation.

\subsection{Diffusion coefficient}\label{sec:diffusion}
\begin{figure}[!t]
    \centering
    \includegraphics[width=0.7\linewidth]{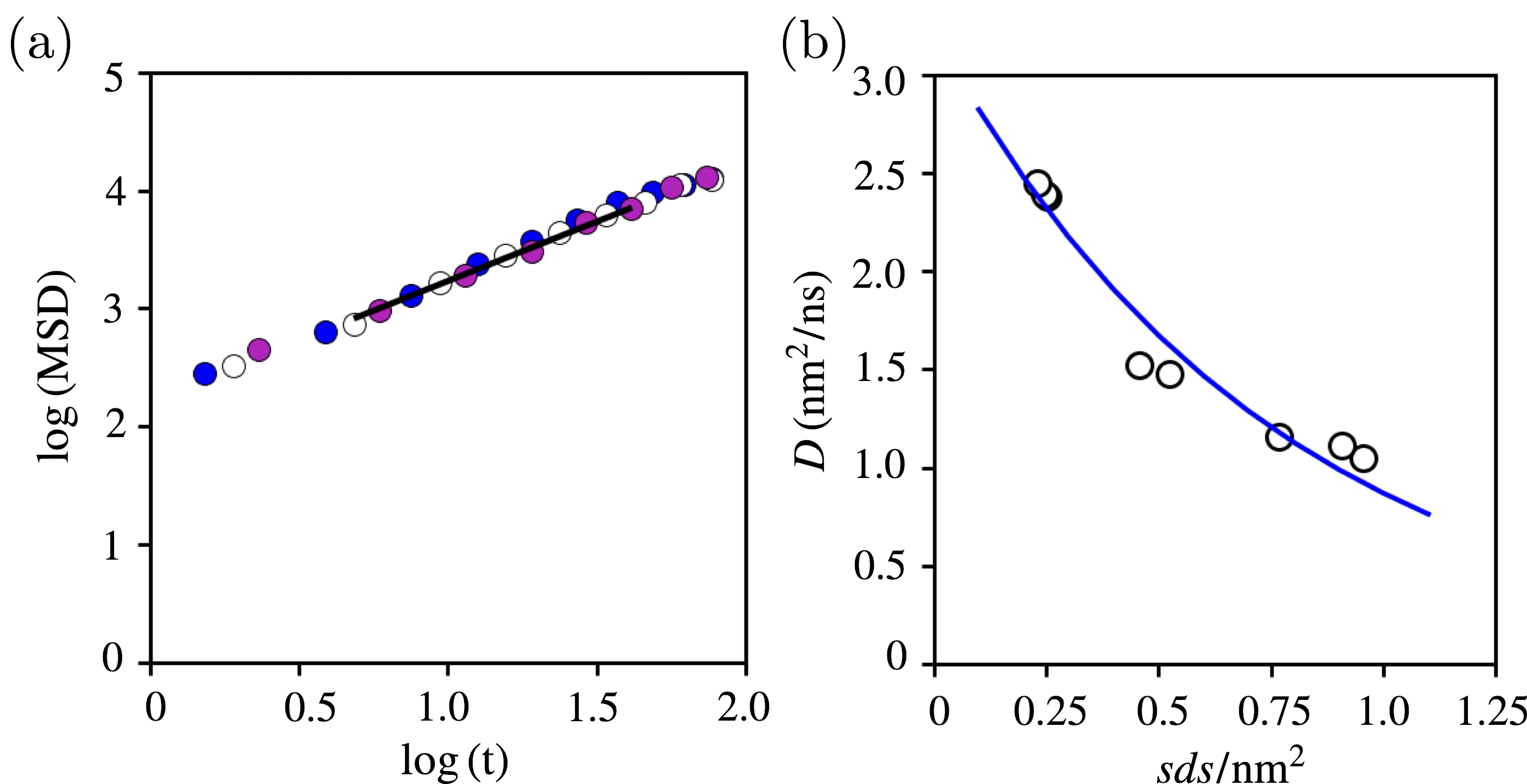}
    \caption{Estimating diffusion coefficient. (a) Mean square displacements of surfactant molecules homogeneously spread over the surface of a film. (b) Diffusion coefficient as a function of surfactant concentration. Blue line denote fitting to the data of the form: $D = \alpha\,e^{-\beta N}$.}
    \label{fig:diff}
\end{figure}

A separate set of simulations were carried out to estimate the diffusion coefficient of the surfactant molecules. Films with homogeneously spread surfactants at the surface were simulated for a range of SDS concentration. 
The trajectories of the surface SDS were tracked and analyzed for the mean squared displacement (MSD). 
Using Einstein's equation, the diffusion coefficient was calculated, see for instance, Supplementary Fig. S\ref{fig:diff} (a) where MSD of the surfactant molecules are plotted from three independent simulations, but for same concentration. 
The average slope gives the diffusion coefficient, $D$. Supplementary Fig. S\ref{fig:diff} (b) shows the variation of $D$ over a range of concentrations. The decreasing trend suggests a concentration-dependent diffusion process, where $D$ decreases with increasing concentration. This is consistent with scenarios where intermolecular interactions become significant at higher concentrations, leading to a reduced mobility of the molecules.

\subsection{Transport equation applied to different initial concentrations} 
\begin{figure}[!t]
    \centering
    \includegraphics[width=1.0\linewidth]{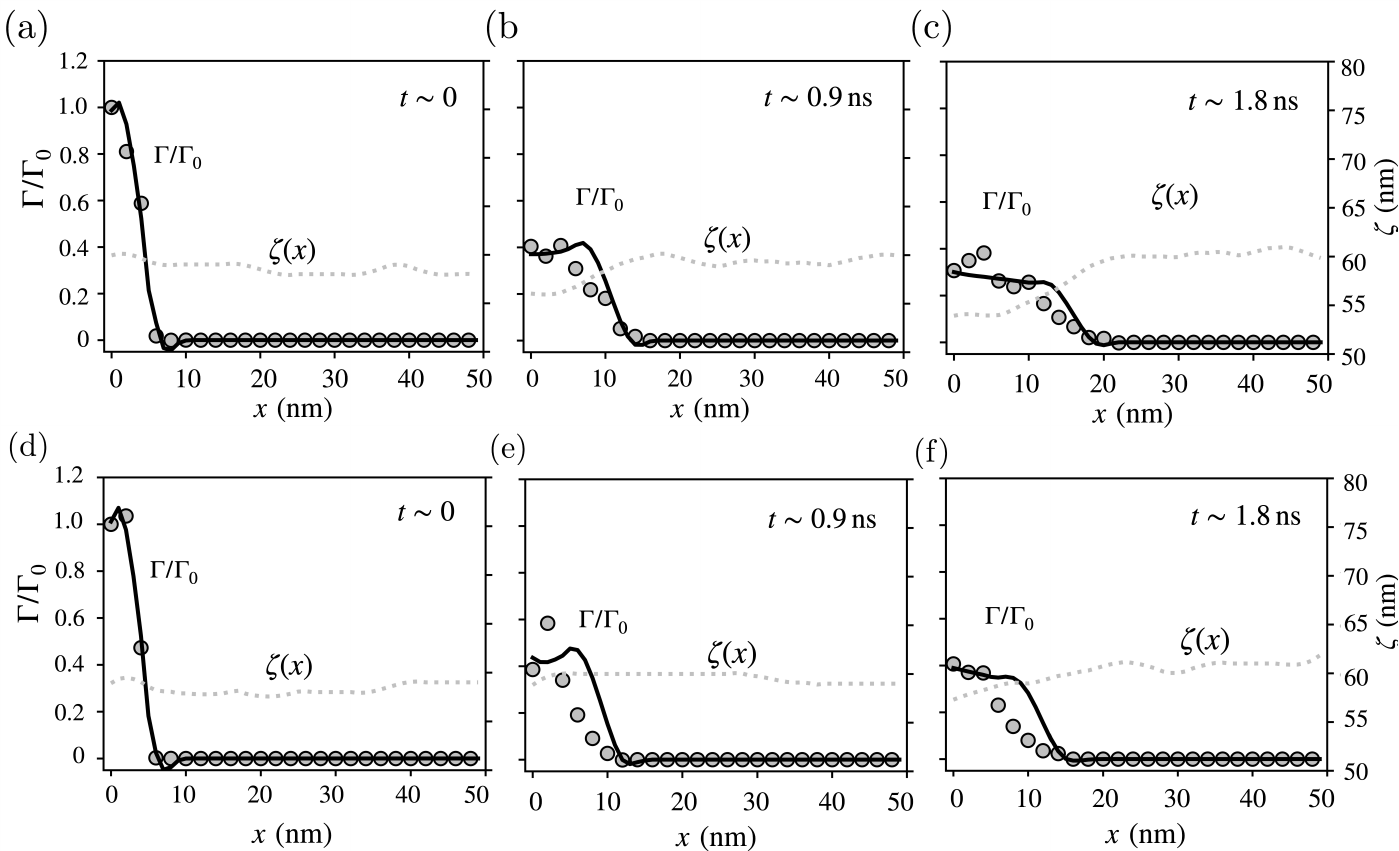}
    \caption{Spatio-temporal variation of surfactant concentration (normalized by initial concentration, the amount of deposited surfactant is $\sim0.75\,\mbox{ng}$ for panels a-c, and $\sim0.2\,\mbox{ng}$ for panels d-f) along the film’s surface. Solid lines denote numerical solution of the transport equation, symbols represent MD data, and the dotted lines denote fitting to the surface.}
    \label{fig:transport}
\end{figure}
In addition to Fig. $2$ in the main text, the agreement between the molecular dynamics simulations, and the transport equation is further evidenced from Fig. S\ref{fig:transport} which shows the time evolution of normalized surfactant concentration along the surface of the film for varied initial amount of deposited surfactant molecules.

\newpage 
\bibliographystyle{unsrtnat}
\bibliography{references}